\documentclass[aps,12pt,notitlepage,final,oneside,onecolumn,nobibnotes,
nofootinbib,superscriptaddress,showpacs,showkeys,centertags,amsmath,amssymb,a4paper]{revtex4}

\usepackage{geometry}
\usepackage{enumitem}
\usepackage{mathtools}
\usepackage[colorlinks,citecolor=blue,urlcolor=blue,linkcolor=blue]{hyperref}

\DeclareMathOperator{\Sp}{Sp}

\DeclareMathOperator{\diag}{diag}
\newcommand{\operP}{\mathcal{P}}
\newcommand{\bra}[1]{\langle #1|}
\newcommand{\ket}[1]{|#1 \rangle}

\newcommand{\journal}[1]{\textit{#1}}

\begin{document}

\title{Mixing of fermions and spectral representation of propagator}

\author{A.E. Kaloshin}
\email{kaloshin@physdep.isu.ru}
\affiliation{Physical Department, Irkutsk State University, K.
  Marx str. 1, 664003, Irkutsk, Russia}
\author{V.P. Lomov}
\email{lomov.vl@icc.ru}
\affiliation{Laboratory 1.2, Institute for System Dynamics
  and Control Theory, RAS, Lermontov str. 134, 664043, Irkutsk,
  Russia}


\begin{abstract}
  We develop the spectral representation of propagator for $n$ mixing fermion
  fields in the case of $\mathsf{P}$-parity violation. The approach based on the
  eigenvalue problem for inverse matrix propagator makes possible to build the
  system of orthogonal projectors and to represent the matrix propagator as a
  sum of poles with positive and negative energies. The procedure of
  multiplicative renormalization in terms of spectral representation is
  investigated and the renormalization matrices are obtained in a closed form
  without the use of perturbation theory. Since in theory with
  $\mathsf{P}$-parity violation the standard spin projectors do not commute with
  the dressed propagator, they should be modified. The developed approach allows
  us to build the modified (dressed) spin projectors for a single fermion and
  for a system of fermions.

\end{abstract}

\maketitle

\section{Introduction}
 
The problem of neutrino oscillations has been in the spotlight since last
decades, both from experimental and theoretical points of view. This phenomenon
is generated by mixing in the neutrinos system, when mass states differ from the
flavor ones. Since quantum field theory is a proper theoretical framework for
describing these effects, the essential efforts were devoted to application of
QFT methods for neutrinos mixing problem
\cite{Gri96,Gri99,Blasone:1995zc,Fujii:2001zv,Giu02,Beu03,Akh10,Nau10,Dvo11,
  Mar12}. What we cited here is only a small part of relevant publications (see
also the references cited therein), which is directly related to problem of
neutrino oscillations in the QFT. The mixing effects also play an essential role
in the quarks system, where radiative corrections lead to modification of the
bare Cabbibo-Kobayashi-Maskawa (CKM) matrix and to necessity to renormalize this
matrix (see,
e.g. Refs. \cite{Donoghue:1979jq,Denner:1990yz,Barroso:2000is,Kniehl:2006rc}). Note
that in the mixing problem there exist some delicate theoretical issues related
with dependence on renormalization scheme, possible gauge dependence and
properties of renormalized mixing matrix \cite{Gam99, Ant06, Dur09}.

In studying of mixing and oscillation phenomena in the QFT the matrix propagator
plays the central role. In recent series of papers \cite{Kni12, Kni14PRL,
  Kni14PR} the properties of dressed matrix propagator in the presence of
$\mathsf{P}$-parity violation were investigated in detail. The dressed
propagator was represented in a closed algebraic form, which satisfies the main
physical requirements and allows to build the renormalized propagator. The pole
scheme of renormalization was investigated and wave-function renormalization
(WFR) matrices were obtained in a closed analytical form without recourse to
perturbation theory.

In the present paper we develop a convenient algebraic construction for
consideration of fermion matrix propagator and mixing effects in the QFT
frameworks. The main feature of suggested construction is that propagator is
represented as a sum of single poles with positive and negative energies. Note,
that it is made in a covariant manner $1/(W\pm m_{i})$ and this is a general
property of considered eigenvalue problem, see e.g. \eqref{freeg} for free
fermion propagator. The obtained very simple expression for WFR matrices
\eqref{eq:renorm-matrices} confirms the old opinion that just $W$ is the natural
variable in fermion case.

Another important feature of the suggested approach is related with spin
properties of the dressed propagator. In theory with $\gamma^{5}$ the usual spin
projectors do not commute with dressed propagator and should be somehow
modified. Standard procedure of Dyson summation (in particular, in
Refs. \cite{Kni12, Kni14PRL, Kni14PR}) does not touch the spin projectors,
having in mind their existence. For the developed here approach the generalized
spin projectors \eqref{spin_n1}, \eqref{spin_n} are the necessary elements of
construction, used to prove the completeness condition.

Technically, the suggested construction is based on so called spectral
representation of an operator (see, e.g. textbook \cite{Messiah:1961qm}). In
this representation the self-adjoint operator $ \hat{A}$ takes the form (in
quantum-mechanical notations):
\begin{equation*}
  \hat{A}=\sum_{i}\lambda_{i}\ket{i}\bra{i}=\sum_{i}\lambda_{i}\Pi_{i},
\end{equation*}
where $\lambda_{i}$ are eigenvalues of the operator, $\ket{i}$ are eigenvectors
\begin{equation*}
  \hat{A}\ket{i}=\lambda_{i}\ket{i}
\end{equation*}
and $\Pi_{i}=\ket{i}\bra{i}$ are corresponding orthogonal projectors
(eigenprojectors). In the case of non-self-adjoint operator the similar
decomposition also exists but to construct it, one needs solutions of both left
and right eigenvalue problems.

If we have $n$ fermion fields with the same quantum numbers, they begin to mix
at loop level even in the case of diagonal mass matrix. In the QFT the main
object of studying is the dressed matrix propagator $G(p)$. To build the
spectral representation of $G(p)$, first of all one needs to solve the
eigenvalue problem for inverse propagator $S(p)$ \footnote{Here $S$ (and $\Pi$
  also) has two sets of indices $S_{\alpha\beta; ij}$, where
  $\alpha,\beta=1,\dots,4$ are the Dirac $\gamma$-matrix indices and
  $i,j=1,\dots,n$ are generation indices. Note that, from the beginning we are
  looking for eigenprojectors instead of eigenvectors (following to
  Ref. \cite{KL12}), to avoid cumbersome intermediate expressions.}
\begin{equation}
  \label{eigenL}
  S \Pi_{i} = \lambda_{i} \Pi_{i} .
\end{equation}
If we have the complete system of orthogonal eigenprojectors\footnote{The
  completeness condition and closely related with it spin projectors are
  discussed in Section \ref{sec:compl-cond-spin}.}
\begin{equation}
  \Pi_{i} \Pi_{k} = \delta_{ik} \Pi_{k} ,\quad
  \sum_{i=1}^{2n}\Pi_{i}=1,
\end{equation}
then we obtain the spectral representation of inverse propagator $S(p)$
\begin{equation}
  \label{inver}
  S(p) = \sum_{i=1}^{2n} \lambda_{i} \Pi_{i}.
\end{equation}

The matrix propagator $G(p)$ is obtained by reversing of \eqref{inver}
\begin{equation}
  \label{prop}
  G(p) = \sum_{i=1}^{2n} \frac{1}{\lambda_{i} }\Pi_{i}.
\end{equation}

If the projectors satisfy the orthogonality property, then the same $\Pi_{i}$
are solutions of two eigenvalue problems: left \eqref{eigenL} and right one
\begin{equation}
  \label{eigenR}
  \Pi_{i} S = \lambda_{i} \Pi_{i} .
\end{equation}

As will be shown later, the representation \eqref{inver} looks very simple and
evident in the case of $\mathsf{P}$-parity conservation and the main technical
problems are related with appearance of $\gamma^{5}$ in vertex and dressed
propagator. In Ref. \cite{KL12} we constructed the representation \eqref{inver}
for a single fermion ($n=1$) in the case of parity violation and investigated
the renormalization procedure. In the present paper we build the spectral
representation for the case of $n$ mixing fermion fields and study the main
properties of this representation.

The paper is organized as follows. In Sec.~\ref{sec:eigenv-probl-matr} we
consider the eigenvalue problem for inverse matrix propagator in theory with
$\mathsf{P}$-parity violation and build the corresponding set of orthogonal
projectors, which are solutions of both left and right eigenvalue problems. In
Sec.~\ref{sec:case-of-cp-conservation} the case of $\mathsf{CP}$-conserving
theory is considered, that is reflected in symmetry of matrix coefficients and
leads to essential simplification of the spectral
representation. Sec.~\ref{sec:compl-cond-spin} is devoted to the completeness
condition for the obtained eigenprojectors, which is equivalent to the existence
of the generalized (dressed) spin projectors with the necessary properties. We
indicate the explicit form of generalized (in theory with $\gamma^{5}$) spin
projectors, which are closely related with the obtained eigenprojectors. In
Sec.~\ref{sec:renorm-scheme} we formulate the multiplicative renormalization
requirements for matrix propagator in terms of the obtained spectral
representation. It gives very simple conditions for the renormalization
constants and allows to write down the answer in a closed form.

\section{Eigenvalue problem for inverse matrix propagator}
\label{sec:eigenv-probl-matr}

\subsection{Preliminary}

In the following it's convenient to use the off-shell $\gamma$-matrix
projectors\footnote{Many people used these off-shell projectors for different
  purposes, the first known for us case is related with the problem of fermion
  Regge poles, see papers of V.N. Gribov and co-authors \cite{Gri63,Gri64}. }
\begin{equation}
  \label{lam}
  \Lambda^{\pm} (p) = \frac{1}{2} \Big( 1\pm \frac{\hat{p}}{W} \Big),
\end{equation}
where $W=\sqrt{p^{2}}$ is in general a complex variable and for positive $p^{2}$
it is the center-of-mass energy. In the study that follows we do not impose any
restrictions on the sign of $p^{2}$. For free inverse propagator
$S_{0}=\hat{p}-m$ these projectors are solutions of eigenvalue problem and free
propagator is represented as
\begin{equation}
  \label{freeg}
  G(p) = \frac{1}{\hat{p}-m} = \frac{1}{W-m} \Lambda^{+}  +
  \frac{1}{-W-m} \Lambda^{-},
\end{equation}
so we obtain a covariant separation of poles with positive and negative
energies.

In the case of parity conservation the eigenprojectors $\Pi_{i}$ are just
$\Lambda^{\pm}$, multiplied by flavor matrix, see \eqref{pcons} below. In the theory
with $\gamma^{5}$ the $\gamma$-matrix projectors $\Lambda^{\pm}$ appear at
intermediate stage of the $\Pi_{i}$ building but they are useful to simplify the
algebra.

In the case of parity violation we introduce the following set of matrices
\begin{equation}
  \label{basis}
  \mathcal{P}_{1}=\Lambda^{+},\quad
  \mathcal{P}_{2}=\Lambda^{-},\quad
  \mathcal{P}_{3}=\Lambda^{+}\gamma^{5},\quad
  \mathcal{P}_{4}=\Lambda^{-}\gamma^{5}
\end{equation}
and use them as a basis to expand the self-energy and propagator. The inverse matrix
propagator may be written as
\begin{equation}
  \label{decomp}
  S(p) = G^{-1}(p)= \sum_{M=1}^{4} \mathcal{P}_{M} S_{M}(W) ,
\end{equation}
where the matrix coefficients $S_{M}$ have the obvious symmetry properties:
\begin{equation}
  S_{2}(W)=S_{1}(-W), \quad
  S_{4}(W)=S_{3}(-W)
\end{equation}
and are calculated as\footnote{Here spur is taken over $\gamma$-matrix indices.}
\begin{equation}
  \begin{aligned}
    S_{1} &= \dfrac{1}{2} \Sp(\mathcal{P}_{1} S),\quad
    & S_{2} &= \dfrac{1}{2} \Sp(\mathcal{P}_{2} S),\\
    S_{3} &= \dfrac{1}{2} \Sp(\mathcal{P}_{4} S),\quad
    & S_{4} &= \dfrac{1}{2} \Sp(\mathcal{P}_{3} S) .
  \end{aligned}
\end{equation}

\begin{itemize}
\item If the parity is conserved, the self-energy
  \begin{equation}
    \begin{multlined}
      \Sigma(p) \equiv A(p^{2})+\hat{p}B(p^{2})=\\[1ex]
      = \mathcal{P}_{1} (A(W^{2})+W B(W^{2}))
      +\mathcal{P}_{2} (A(W^{2})-W B(W^{2}))
    \end{multlined}
  \end{equation}
  contains only two terms in the decomposition \eqref{decomp}. In this case the
  eigenvalue problem \eqref{eigenL} is reduced to eigenvalue problem for $n
  \times n$ matrices $S_{1,2}$
  \begin{equation}
    \label{pcon}
    \begin{split}
      S_{1}\pi_{1} &\equiv (A(W^{2})+WB(W^{2})) \pi_{1} = \lambda \pi_{1}, \\
      S_{2}\pi_{2} &\equiv (A(W^{2})-WB(W^{2})) \pi_{2} = \lambda \pi_{2}
    \end{split}
  \end{equation}
  and eigenprojectors $\Pi_{i}$ take the factorized form
  \begin{equation}
    \label{pcons}
    \begin{split}
      \Pi_{i} &= \Lambda^{+} \pi_{1}^{(i)} , \quad i=1,\dots,n\,, \\
      \Pi_{i} &= \Lambda^{-} \pi_{2}^{(i)} , \quad i=n+1,\dots,2n
    \end{split}
  \end{equation}
  for positive and negative energy poles respectively.
\item If $\mathsf{P}$-parity is violated, the spectral representation
  \eqref{inver} for inverse propagator becomes less evident. For single fermion
  ($n=1$ in the above) it was built and investigated in Ref. \cite{KL12}. The
  eigenvalues $\lambda_{1,2}(W)$ are defined by the characteristic equation
  \begin{equation}
    \label{eq:ce-ev}
    \lambda^{2}-\lambda(S_{1}+S_{2})+(S_{1}S_{2}-S_{3}S_{4})=0,
  \end{equation}
  where the numbers $S_{i}(W)$ are coefficients in the decomposition
  \eqref{decomp}. The eigenprojectors in general case are
  \begin{equation}
    \label{eq:ip-projs}
    \begin{split}
      \Pi_{1} &= \frac{1}{\lambda_{2}-\lambda_{1}}
      \Bigl((S_{2}-\lambda_{1})\operP_{1}+(S_{1}-\lambda_{1})\operP_{2}
      -S_{3}\operP_{3}-S_{4}\operP_{4}\Bigr),\\
      \Pi_{2} &= \frac{1}{\lambda_{1}-\lambda_{2}}
      \Bigl((S_{2}-\lambda_{2})\operP_{1}+(S_{1}-\lambda_{2})\operP_{2}
      -S_{3}\operP_{3}-S_{4}\operP_{4}\Bigr).
    \end{split}
  \end{equation}
  Finally, note that if to use the $\gamma$-matrix basis for inverse propagator
  \begin{equation}
    S=a+\hat{n}b +\gamma^{5} c + \hat{n}\gamma^{5} d
    =a+\hat{n}(b+\hat{n}\gamma^{5} c + \gamma^{5} d) ,
  \end{equation}
  then the eigenprojectors \eqref{eq:ip-projs} may be rewritten in the very
  simple form
  \begin{equation}
    \Pi_{1,2}=\dfrac{1}{2} \bigg( 1 \pm \hat{n}\cdot
    \dfrac{b+\hat{n}\gamma^{5} c + \gamma^{5} d}{\sqrt{b^{2}+c^{2}-d^{2}}}
    \bigg) ,
  \end{equation}
  where $n^{\mu}=p^{\mu}/W$ is the unit vector.
\end{itemize}

\subsection{Left eigenvalue problem}

Let us consider the mixing problem for $n$ fermion fields in the theory with
parity violation. The inverse propagator is defined by decomposition
\eqref{decomp} with arbitrary matrix coefficients $S_{M}(W)$. Following
Ref. \cite{KL12}, we solve the eigenvalue problem
\begin{equation}
  \label{eigenLL}
  S \Pi = \lambda \Pi
\end{equation}
in matrix form, i.e. from the beginning we are looking for eigenprojectors $\Pi$
instead of eigenvectors. The sought-for eigenprojectors may also be written as
decomposition \eqref{decomp}
\begin{equation}
  \label{leftPi}
  \Pi= \sum_{M=1}^{4} \mathcal{P}_{M} A_{M} ,
\end{equation}
with matrix $n\times n$ coefficients $A_{M}(W)$. Due to simple multiplicative
properties of the basis \eqref{basis}, it's easy to reduce the eigenvalue
problem \eqref{eigenLL} to the following set of linear equations for unknown
matrices $A_{M}$
\begin{equation}
  \begin{split}
    (S_{1}-\lambda)A_{1} + S_{3} A_{4} &= 0, \\
    (S_{2}-\lambda)A_{2} + S_{4} A_{3} &= 0, \\
    (S_{1}-\lambda)A_{3} + S_{3} A_{2} &= 0, \\
    (S_{2}-\lambda)A_{4} + S_{4} A_{1} &= 0.
  \end{split}
\end{equation}

In fact we have two separated subsystems for unknown $A_{1}$, $A_{4}$ and
$A_{2}$, $A_{3}$, so it's convenient to express $A_{3}$, $ A_{4}$ by
\begin{equation}
  A_{3}=-S_{4}^{-1}(S_{2}-\lambda)A_{2}, \quad
  A_{4}=-S_{3}^{-1}(S_{1}-\lambda)A_{1}
\end{equation}
and to obtain the homogeneous equations for $n \times n$ matrices $A_{1}$,
$A_{2}$
\begin{equation}
  \label{homo}
  \begin{split}
    \hat{O}A_{1} &\equiv
    [(S_{2}-\lambda)S_{3}^{-1}(S_{1}-\lambda)-S_{4}] A_{1}=0, \\
    \hat{O}^{\prime} A_{2} &\equiv
    [(S_{1}-\lambda)S_{4}^{-1}(S_{2}-\lambda)-S_{3}] A_{2}=0 .
  \end{split}
\end{equation}
Here we introduced the short notations $\hat{O}$, $\hat{O}^{\prime}$ for
appeared $\lambda$-dependent operators.  One can see that matrices $\hat{O}$,
$\hat{O}^{\prime}$ are connected by similarity relationship
\begin{equation}
  \label{OO}
  \hat{O}^{\prime}=(S_{1}-\lambda)S_{4}^{-1}\cdot \hat{O} \cdot
  (S_{1}-\lambda)^{-1} S_{3}
  = S_{3}(S_{2}-\lambda)^{-1}\cdot \hat{O} \cdot
  S_{4}^{-1} (S_{2}-\lambda),
\end{equation}
so equations \eqref{homo} give the same characteristic equation for $\lambda$
\begin{equation}
  \label{det}
  \det [(S_{2}-\lambda)S_{3}^{-1}(S_{1}-\lambda)-S_{4}] = 0.
\end{equation}
In the absence of degeneration this equation gives $2n$ different eigenvalues
$\lambda_{i}(W)$.

Thus, the matrix solution of left eigenvalue problem \eqref{eigenLL} may be
written as
\begin{equation}
  \label{sleft}
  \Pi^{i} = \mathcal{P}_{1} A_{1}^{i} + \mathcal{P}_{2} A_{2}^{i}
          - \mathcal{P}_{3} S_{4}^{-1}(S_{2}-\lambda_{i})A_{2}^{i} -
            \mathcal{P}_{4} S_{3}^{-1}(S_{1}-\lambda_{i})A_{1}^{i} ,
\end{equation}
where matrices $A_{1}^{i}$, $A_{2}^{i}$ are solutions of equations
\begin{equation}
  \label{homoi}
  \begin{split}
    \hat{O}_{i} A_{1}^{i} &\equiv \hat{O}(\lambda=\lambda_{i}) A_{1}^{i}=0,\\
    \hat{O}_{i}^{\prime} A_{2}^{i} &\equiv
    \hat{O}^{\prime}(\lambda=\lambda_{i}) A_{2}^{i} =0
  \end{split}
\end{equation}
and eigenvalues $\lambda_{i}(W)$ are defined by equation \eqref{det}.

\subsection{Right eigenvalue problem}

It was noted in the above that orthogonal projectors should satisfy both left
and right eigenvalue problems. So as the next step consider the right eigenvalue
problem for inverse propagator
\begin{equation}
  \label{eigenRR}
  \Pi_{R} S= \lambda \Pi_{R} .
\end{equation}
We can look for the right eigenprojectors $\Pi_{R}$ in the same form
\eqref{leftPi} with matrix coefficients $B_{M}$. Similar calculations give the
matrix solution of the right problem
\begin{equation}
  \label{sright}
  \Pi^{i}_{R} = \mathcal{P}_{1} B_{1}^{i} + \mathcal{P}_{2} B_{2}^{i}
  - \mathcal{P}_{3} B_{1}^{i} S_{3} (S_{2}-\lambda_{i})^{-1} -
  \mathcal{P}_{4} B_{2}^{i} S_{4} (S_{1}-\lambda_{i})^{-1} ,
\end{equation}
where $B_{1}^{i}$, $B_{2}^{i}$ are solutions of the right homogeneous equations
\begin{equation}
  \label{homR}
  B_{1}^{i} \hat{O}_{i} ^{\prime} = 0, \qquad
  B_{2}^{i} \hat{O}_{i} = 0
\end{equation}
and eigenvalues $\lambda_{i}(W)$ are defined by the same equation \eqref{det}.

\subsection{Left and right problems together}

Let us require the ``matrix'' $\Pi$ to be solution of both left and right
eigenvalue problems. It means that expressions \eqref{sleft}, \eqref{sright}
should coincide with each other.

First of all $B_{1}^{i}=A_{1}^{i}$, $B_{2}^{i}=A_{2}^{i}$, as is seen from
$\mathcal{P}_{1}$, $\mathcal{P}_{2}$ terms. Coefficients at $\mathcal{P}_{3}$,
$\mathcal{P}_{4}$ give two relations between $A_{1}$ and $A_{2}$
\begin{equation}
  \label{A1A2}
  \begin{split}
    A_{2}^{i} &=  S_{3}^{-1} (S_{1}-\lambda_{i})\cdot
    A_{1}^{i} \cdot S_{3} (S_{2}-\lambda_{i})^{-1},\\
    A_{2}^{i} &= (S_{2}-\lambda_{i})^{-1} S_{4}\cdot
    A_{1}^{i} \cdot (S_{1}-\lambda_{i}) S_{4}^{-1} .
  \end{split}
\end{equation}

Now the matrices $A_{1}$, $A_{2}$ should satisfy both left and right homogeneous
equations
\begin{equation}
  \label{homoLR}
  \begin{aligned}
    \hat{O}_{i} A_{1}^{i} &= 0, \quad & A_{1}^{i} \hat{O}_{i}^{\prime} &= 0 ,\\
    \hat{O}_{i}^{\prime} A_{2}^{i} &= 0, \quad & A_{2}^{i} \hat{O}_{i} &= 0 ,
  \end{aligned}
\end{equation}
where the matrices $\hat{O}_{i}$, $\hat{O}_{i}^{\prime}$ are defined by
\eqref{homo}.

Note that homogeneous equations for $A_{1}$ lead to the following equalities
\begin{equation}
  \begin{split}
    S_{3}^{-1}(S_{1}-\lambda_{i})\cdot A_{1}^{i}
    &= (S_{2}-\lambda_{i})^{-1} S_{4} \cdot A_{1}^{i} ,\\
    A_{1}^{i} \cdot (S_{1}-\lambda_{i}) S_{4}^{-1}
    &= A_{1}^{i} \cdot S_{3} (S_{2}-\lambda_{i})^{-1} ,
  \end{split}
\end{equation}
so one can see that two relations \eqref{A1A2} actually coincide. Moreover, one
can convince yourself that equations for $A_{2}^{i}$ \eqref{homoLR} are
consequence of relation \eqref{A1A2} and equations for $A_{1}^{i}$. Therefore,
it is sufficient to require the left and right homogeneous equations for
$A_{1}^{i}$ (first line in \eqref{homoLR}) and connection between $A_{2}^{i}$
and $A_{1}^{i}$ (one of \eqref{A1A2}).

At last, note that the matrix $A_{1}^{i}$ has zeroth determinant and may be
represented in the split form
\begin{equation}
  \label{A1psi}
  A_{1}^{i} = \psi_{i} (\tilde{\psi}_{i})^{\text{T}} ,
\end{equation}
where vectors $\psi_{i}$, $\tilde{\psi}_{i}$ (columns) are solutions of
homogeneous equations
\begin{equation}
  \label{eq_psi}
  \hat{O}_{i} \psi_{i} = 0,\quad (\tilde{\psi}_{i})^{\text{T}}
  \hat{O}_{i}^{\prime} = 0 \qquad
  \Big( \text{or } (\hat{O}_{i}^{\prime})^{\text{T}} \tilde{\psi}_{i} =0 \Big).
\end{equation}
Then solution of both left and right eigenvalue problems may be represented as
\begin{multline}
  \label{eig_LR}
  \Pi_{i} = \mathcal{P}_{1} \psi_{i} (\tilde{\psi}_{i})^{\text{T}} +
  \mathcal{P}_{2} S_{3}^{-1} (S_{1} - \lambda_{i}) \psi_{i}
  (\tilde{\psi}_{i})^{\text{T}} (S_{1} - \lambda_{i}) S_{4}^{-1} - \\ 
  -\mathcal{P}_{3} \psi_{i} (\tilde{\psi}_{i})^{\text{T}}(S_{1} -
  \lambda_{i}) S_{4}^{-1} - \mathcal{P}_{4} S_{3}^{-1} (S_{1} -
  \lambda_{i}) \psi_{i} (\tilde{\psi}_{i})^{\text{T}} .
\end{multline}
For short notations, it is convenient to introduce the vectors $\phi_{i}$,
$\tilde{\phi}_{i}$ as
\begin{equation}
  \label{def_phi}
  \phi_{i} = S_{3}^{-1} (S_{1} - \lambda_{i}) \psi_{i} ,\quad
  (\tilde{\phi}_{i})^{\text{T}} = (\tilde{\psi}_{i})^{\text{T}}
  (S_{1} - \lambda_{i}) S_{4}^{-1} .
\end{equation}

In these terms the ``matrix'' $\Pi_{i}$, which is a solution of both left and
right eigenvalue problems, takes very elegant form
\begin{equation}
  \label{L_R}
  \Pi_{i} = \mathcal{P}_{1}\cdot \psi_{i} (\tilde{\psi}_{i})^{\text{T}} +
  \mathcal{P}_{2}\cdot \phi_{i} (\tilde{\phi}_{i})^{\text{T}} -
  \mathcal{P}_{3}\cdot \psi_{i} (\tilde{\phi}_{i})^{\text{T}} -
  \mathcal{P}_{4}\cdot \phi_{i} (\tilde{\psi}_{i})^{\text{T}} .
\end{equation}

Recall, that the auxiliary vectors $\phi_{i}$, $\tilde{\phi}_{i}$ also satisfy
the following homogeneous equations (consequence of definition)
\begin{equation}
  \label{eq_phi}
  \hat{O}_{i}^{\prime} \phi_{i} = 0,\quad
  (\tilde{\phi}_{i})^{\text{T}} \hat{O}_{i} = 0 .
\end{equation}

\subsection{Eigenprojectors}

So we have $\Pi_{i}$ \eqref{L_R} --- solutions of both left and right eigenvalue
problems. Let us require these ``matrices'' (with two sets of indices) $\Pi_{i}$
to be orthogonal projectors
\begin{equation}
  \Pi_{i} \Pi_{k} = \delta_{ik}\Pi_{k} .
\end{equation}
It gives four equations if to use the decomposition \eqref{decomp}
\begin{equation}
  \label{proj}
  \begin{split}
    \psi_{i} \Big[(\tilde{\psi}_{i})^{\text{T}}  \psi_{k}  +
    (\tilde{\phi}_{i})^{\text{T}} \phi_{k} - \delta_{ik}\Big]
    (\tilde{\psi}_{k})^{\text{T}} &= 0 , \\
    \phi_{i} \Big[(\tilde{\psi}_{i})^{\text{T}}  \psi_{k}  +
    (\tilde{\phi}_{i})^{\text{T}} \phi_{k} - \delta_{ik}\Big]
    (\tilde{\psi}_{k})^{\text{T}} &= 0 , \\
    \psi_{i} \Big[(\tilde{\psi}_{i})^{\text{T}}  \psi_{k}  +
    (\tilde{\phi}_{i})^{\text{T}} \phi_{k} - \delta_{ik}\Big]
    (\tilde{\phi}_{k})^{\text{T}} &= 0 , \\
    \phi_{i} \Big[(\tilde{\psi}_{i})^{\text{T}} \psi_{k} +
    (\tilde{\phi}_{i})^{\text{T}} \phi_{k} - \delta_{ik}\Big]
    (\tilde{\phi}_{k})^{\text{T}} &= 0 ,
  \end{split}
\end{equation}
which are equivalent to the orthonormality condition for vectors involved in
\eqref{L_R}
\begin{equation}
  \label{ortho}
  (\tilde{\psi}_{i})^{\text{T}}  \psi_{k}  +
  (\tilde{\phi}_{i})^{\text{T}} \phi_{k} = \delta_{ik} .
\end{equation}

\begin{itemize}
\item If $i \neq k$ the condition \eqref{ortho} is consequence of equation on
  $\psi_{k}$ and $(\tilde{\psi}_{i})^{\text{T}}$. To see it, let us rewrite
  \eqref{ortho} in terms of the vectors $\psi_{i}$ and $\tilde{\phi}_{i}$:
  \begin{equation}
    \label{ortho1}
    (\tilde{\phi}_{i})^{\text{T}} \Big[  (S_{2} - \lambda_{i}) S_{3}^{-1}+
    S_{3}^{-1} (S_{1} - \lambda_{k}) \Big] \psi_{k}  = \delta_{ik} .
  \end{equation}

  Now let us write down the homogeneous equations for $\psi_{k}$ and
  $\tilde{\phi}_{i}$
  \begin{equation}
    \begin{split}
      0 &=\hat{O}_{k} \psi_{k} =
      \Big[S_{3}^{-1} \lambda_{k} ^{2} - \lambda_{k}
      (S_{2} S_{3}^{-1} + S_{3}^{-1}S_{1})
      + S_{2} S_{3}^{-1} S_{1} - S_{4} \Big] \psi_{k} ,\\
      0 &= (\tilde{\phi}_{i})^{\text{T}} \hat{O}_{i} =
      (\tilde{\phi}_{i})^{\text{T}} \Big[S_{3}^{-1} \lambda_{i} ^{2} -
      \lambda_{i} (S_{2} S_{3}^{-1} + S_{3}^{-1}S_{1})
      + S_{2} S_{3}^{-1} S_{1} - S_{4} \Big] .
    \end{split}
  \end{equation}
  Multiplying first of these equations by $(\tilde{\phi}_{i})^{\text{T}}$ from
  the left, second one by $\psi_{k}$ from the right, and subtracting one
  equation from another, we obtain
  \begin{equation}
    (\lambda_{k} - \lambda_{i})
    \cdot(\tilde{\phi}_{i})^{\text{T}}
    \Big[ (S_{2} - \lambda_{i}) S_{3}^{-1} +
    S_{3}^{-1} (S_{1} - \lambda_{k}) \Big]\psi_{k}  = 0
 \end{equation}
 and at $\lambda_{i} \neq \lambda_{k}$ it gives the condition \eqref{ortho}.
\item At $i=k$ equation \eqref{ortho} defines the normalization (with weight) of
  the vector $\psi_{i}$ in respect to $\tilde{\psi}_{i}$.
\end{itemize}

\section{Case of $\mathsf{CP}$ conservation}
\label{sec:case-of-cp-conservation}

In the case of $\mathsf{CP}$ conservation, the matrix $n\times n$ coefficients
of the self-energy contribution
\begin{equation}
  \label{sigma}
  \Sigma(p) = \sum_{M=1}^{4} \mathcal{P}_{M} \Sigma_{M}(W)
  = A(p^{2}) + \hat{p} B(p^{2}) + \gamma^{5} C(p^{2}) +
  \hat{p} \gamma^{5} D(p^{2})
\end{equation}
have the following symmetry properties (see, e.g. Ref. \cite{Kni08})
\begin{equation}
  \label{sim}
  A^{\text{T}} = A, \quad B^{\text{T}}=B,\quad D^{\text{T}} = D, \quad
  C^{\text{T}} =- C,
\end{equation}
which are equivalent to
\begin{equation}
  \label{sim1}
  (\Sigma_{1,2})^{\text{T}} = \Sigma_{1,2}, \quad
  (\Sigma_{3})^{\text{T}}=-\Sigma_{4} .
\end{equation}
Since the inverse propagator $S(p)$ has the same symmetry properties
\eqref{sim1}, it connects matrices $\hat{O}$ and $\hat{O}^{\prime}$ \eqref{homo}
\begin{equation}
  \label{OOCP}
  \hat{O}^{\prime} = - (\hat{O})^{\text{T}} .
\end{equation}

Eigenprojectors have the form \eqref{L_R} but now two equations \eqref{eq_psi}
coincide
\begin{equation}
  \label{eq_psiCP}
  \hat{O}_{i} \psi_{i} = 0,\quad \hat{O}_{i} \tilde{\psi}_{i} = 0 .
\end{equation}

Then, in the absence of degeneration, we have
$\tilde{\psi}_{i} = c_{i} \psi_{i}$ and the coefficient $c_{i}$ may be absorbed
by redefinition of vector. From the limiting case of parity conservation (see
Sec.~\ref{sec:case-parity-cons}) it follows that $c_{i}$ should have different
signs for solution with positive and negative energies. So, the most convenient
choice is $\tilde{\psi}_{i} = \varepsilon_{i} \psi_{i}$, where
$\varepsilon_{i} =\pm 1$ is the sign of energy.

So, the eigenprojectors \eqref{L_R} in the case of $\mathsf{CP}$ conservation
take the form
\begin{equation}
  \label{L_R_CP}
  \Pi_{i} = \varepsilon_{i} \big(\mathcal{P}_{1}\cdot \psi_{i} ({\psi}_{i})^{\text{T}} -
  \mathcal{P}_{2}\cdot \phi_{i} ({\phi}_{i})^{\text{T}} +
  \mathcal{P}_{3}\cdot \psi_{i} ({\phi}_{i})^{\text{T}} -
  \mathcal{P}_{4}\cdot \phi_{i} ({\psi}_{i})^{\text{T}}\big) 
\end{equation}
and the vector $\phi_{i}$ is related to $\psi_{i}$ by
\begin{equation}
  \label{cp_rel}
  \phi_{i} = S_{3}^{-1} (S_{1} - \lambda_{i}) \psi_{i},\quad
  \text{or}\quad
  ({\phi}_{i})^{\text{T}} =- ({\psi}_{i})^{\text{T}}
  (S_{1} - \lambda_{i}) S_{4}^{-1} .
\end{equation}

In the case of $\mathsf{CP}$ conservation, we need to solve the homogeneous
equation for vector $\psi_{i}$ for every $\lambda_{i}$
\begin{equation}
  \label{cp_hom}
  \hat{O}_{i} \psi_{i} =
  \big[ (S_{2} - \lambda_{i})S_{3}^{-1}(S_{1}-\lambda_{i}) - S_{4} \big]
  \psi_{i} = 0 , \quad
  i=1, \dots , 2n
\end{equation}
and to calculate $\phi_{i}$ according to \eqref{cp_rel}. Note that $\phi_{i}$
satisfies the homogeneous equation (consequence of \eqref{cp_hom},
\eqref{cp_rel})
\begin{equation}
  \label{cp_phi}
  \hat{O}_{i}^{T} \phi_{i} = -
  \big[ (S_{1} - \lambda_{i})S_{4}^{-1}(S_{2}-\lambda_{i}) - S_{3} \big]
  \phi_{i} = 0 .
\end{equation}
The orthonormality condition $\Pi_{i} \Pi_{k} = \delta_{ik}\Pi_{k}$ leads to
simple property of vectors
\begin{equation}
  \label{ortho11}
  \varepsilon_{i} \big(
  ({\psi}_{i})^{\text{T}}  \psi_{k}  - ({\phi}_{i})^{\text{T}} \phi_{k}
  \big) = \delta_{ik} .
\end{equation}
As it was shown before, this is not a new requirement: at $i\neq k$ it follows
from homogeneous equation and at $i= k$ it defines normalization of vectors
$\psi_{i}$.

But to keep the solutions with positive and negative energies on equal footing
(see Sec.~\ref{sec:case-parity-cons}) one should proceed in a different way.
\begin{enumerate}[label=\alph*)]
\item For the positive energy solution ($i=1, \dots , n$) we solve the equation
  for vector $\psi_{i}$ \eqref{cp_hom} and after it calculate $\phi_{i}$
  according to \eqref{cp_rel}.
\item For the negative energy solution ($i=n+1, \dots , 2n$) we find vector
  $\phi_{i}$ from the equation \eqref{cp_phi}. Then we can calculate the vector
  $\psi_{i}$ from relation\footnote{In fact one can avoid the solution of
    equation \eqref{cp_phi} due to $W\to -W$ replacement --- see, e.g. a
    particular case \eqref{a12}.} \eqref{cp_rel}
  \begin{equation}
    \psi_{i} = S_{4}^{-1}(S_{2}-\lambda_{i}) \phi_{i}.
  \end{equation}
\end{enumerate}

\subsection{Case of parity conservation}
\label{sec:case-parity-cons}

Let us consider a particular case of the spectral representation of propagator,
when parity is conserved\footnote{We suppose that the mixing fermion fields
  $\Psi_{1}$, $\Psi_{2}$ have the same parity (quarks or leptons). If they have
  the opposite parities (baryon fields in effective theories), the self-energy
  contains $\gamma^{5}$ in case of parity conservation and the dressed matrix
  propagator has absolutely different form, see Refs. \cite{Kaloshin:2010jj,
    Kaloshin:2013dha}.}. It allows to clarify some details of general
construction.

In this case the eigenprojectors $\Pi_{i}$
\begin{equation}
  \label{a1}
  S \Pi_{i} =\lambda_{i} \Pi_{i} , \quad \quad
  \Pi_{i} S =\lambda_{i} \Pi_{i}
\end{equation}
take the factorized form, see \eqref{pcons}. Here $n \times n$ matrices
$\pi_{i}$ satisfy the homogeneous equations \eqref{pcon}
\begin{equation}
  \label{a2}
  \begin{split}
    S_{1}\pi_{1} &= \lambda \pi_{1}, \\
    S_{2}\pi_{2} &= \lambda \pi_{2}
  \end{split}
\end{equation}
and also the right equations (see \eqref{eigenR})
\begin{equation}
  \label{a3}
  \begin{split}
    \pi_{1} S_{1} &= \lambda \pi_{1}, \\
    \pi_{2} S_{2} &= \lambda \pi_{2} .
  \end{split}
\end{equation}

It's known that the eigenvalues of left and right problems coincide and since
the matrices $S_{1}(W)$, $S_{2}(W)$ are symmetric ones, the solutions (vectors)
of both left and right eigenvalue probems also coincide. So the matrices
$\pi_{i}$ may be represented in a split form.
\begin{itemize}
\item Projectors which correspond to positive energy poles are given by
  \begin{equation}
    \label{a4}
    \Pi_{i} = \Lambda^{+}(p) \psi_{i} \psi_{i}^{\text{T}}, \qquad
    i=1, \dots, n.  
  \end{equation}
  Vectors $\psi_{i}$ satisfy the eigenvalue equation
  \begin{equation}
    \label{a5}
    S_{1} \psi_{i} = \lambda_{i} \psi_{i}, \qquad
    i=1, \dots, n  ,
  \end{equation}
  where $\lambda_{i}$ are solutions of characteristic equation
  \begin{equation}
    \label{a6}
    \det (S_{1}(W) - \lambda I_{n}) = 0 .
  \end{equation}
\item Projectors onto the negative energy poles are
  \begin{equation}
    \label{a7}
    \Pi_{i} = \Lambda^{-}(p) \phi_{i} \phi_{i}^{\text{T}}, \qquad
    i=n+1, \dots, 2n.
  \end{equation}
  Equation for vectors $\phi_{i}$ is
  \begin{equation}
    \label{a8}
    S_{2} \phi_{i} = \lambda_{i} \phi_{i}, \qquad
    i=n+1, \dots, 2n.
  \end{equation}
  Corresponding characteristic equation is
  \begin{equation}
    \label{a9}
    \det (S_{2}(W) - \lambda I_{n}) = 0.
  \end{equation}
\end{itemize}
Since the matrix coefficients of propagator decomposition are related by
\begin{equation}
  \label{a10}
  S_{2}(W) = S_{1}(-W) ,
\end{equation}
it is sufficient to solve the equations \eqref{a5}, \eqref{a6}, after which the
solutions of \eqref{a8}, \eqref{a9} may be obtained by the replacement
$W \to -W$.

It is convenient to number the eigenvalues in such a way that $\lambda_{i}(W)$
and $\lambda_{i+n}(W)$ would have zeroes at the points $W=m_{i}$ and $W=-m_{i}$
respectively. To this end one should require the relation between solutions of
characteristic equations \eqref{a6}, \eqref{a9}
\begin{equation}
  \label{a11}
  \lambda_{i+n}(W) =  \lambda_{i}(-W), \qquad
  i=1, \dots, n.
\end{equation}

If so, solutions of \eqref{a8} may be obtained from the solutions of \eqref{a5}
(in the absence of degeneration)
\begin{equation}
  \label{a12}
  \phi_{i+n}(W) =  \psi_{i}(-W), \qquad
  i=1, \dots, n  .
\end{equation}

Looking at the homogenious equation \eqref{a5}, one can see that due to symmetry
of the matrix $S_{1}^{\text{T}} = S_{1}$, solutions corresponding to different
$\lambda$ are orthogonal to each other
\begin{equation}
  \label{a13}
  (\lambda_{i} - \lambda_{k}) \psi_{i}^{\text{T}} \psi_{k} = 0,\qquad
  i,k = 1, \dots, n.
\end{equation}

So one can choose them to be orthonormal
\begin{equation}
  \label{a14}
  \psi_{i}^{\text{T}} \psi_{k} = \delta_{ik}, \qquad
  i,k = 1, \dots, n,
\end{equation}
which leads to completeness condition for $n\times n$ matrix
\begin{equation}
  \label{a15}
  \sum_{i=1}^{n} \psi_{i} \psi_{i}^{\text{T}} = I_{n} , 
\end{equation}
where $I_{n}$ is unit matrix of dimension $n$. Then the operators $\Pi_{i}$
\eqref{a4}, \eqref{a7} are the system of $2n$ orthogonal projectors
\begin{equation}
  \label{a16}
  \Pi_{i}   \Pi_{k} =  \delta_{ik} \Pi_{i} , \qquad
  i=1, \dots, 2n  .
\end{equation}

The case of parity conservation described here may also be obtained from the
general spectral representation, for definiteness let us say about the case of
$\mathsf{CP}$-conservation (see Sec.~\ref{sec:case-of-cp-conservation}). In this
general construction one should ``turn off'' the parity violation.

Recall, that in general case the vectors $\psi_{i}$ satisfy the homogenious
equation \eqref{cp_hom}
\begin{equation}
  \label{a17}
  \Big[ (S_{2} - \lambda_{i}) S_{3}^{-1}(S_{1} - \lambda_{i}) -
         S_{4} \Big] \psi_{i} = 0 .
\end{equation}
To return to parity conservation, we should take the limit $S_{3} \to 0$,
$S_{4} \to 0$ in this equation. We see that the characteristic equation in this
limit splits into two factors
\begin{equation}
  \label{a18}
  \det (S_{1} - \lambda)=0, \qquad
  \det (S_{2} - \lambda)=0 .
\end{equation}

For solutions with positive energy (we number them from $1$ to $n$)
\begin{equation}
  \label{a19}
  (S_{1} - \lambda_{i}) \psi_{i} = 0,\qquad
  i = 1, \dots, n  
\end{equation}
and according to relation \eqref{cp_rel} vector $\phi_{i}=0$.

On the contrary, for solutions with negative energy one should solve equation
for $\phi_{i}$ \eqref{cp_phi}
\begin{equation}
  \label{a20}
  (S_{2} - \lambda_{i}) \phi_{i} = 0, \qquad
  i = n+1, \dots, 2n  
\end{equation}
and then to calculate $\psi_{i}$. According to relation \eqref{cp_rel} we get
$\psi_{i} = 0$.

As was noted in the above, the property of eigevalues
$\lambda_{i+n}(W) = \lambda_{i}(-W)$ allows to avoid solving the equation
\eqref{a20} and to use instead the $W\to - W$ replacement
\begin{equation}
  \label{a21}
  \phi_{i+n}(W) = \psi_{i}(-W), \qquad
  i = 1, \dots, n .
\end{equation}

For illustration, let us take a look at particular case of mixing of two fermion
fields in theory with parity conservation.  In this case the energy projection
operators $\Pi_{i}$ $(i=1,\dots,4)$ have the form \eqref{a4}, \eqref{a7}. Let us
write down the parametrization for solutions of eq. \eqref{a5}.
\begin{equation}
  \label{c1}
  \psi_{1}=
  \begin{pmatrix}
    \cos{\theta}\\
    \sin{\theta}
  \end{pmatrix},\quad
  \psi_{2}=
  \begin{pmatrix}
    -\sin{\theta}\\
    \cos{\theta}
  \end{pmatrix},\qquad
  \phi_{1}=\phi_{2}=0 ,
\end{equation}
where we introduced some function $\theta(W)$. We suppose that the self-energy
is real, in this case the solutions $\psi_{1}$, $\psi_{2}$ are also
orthogonal to each other.

Then, according to \eqref{a21} vectors for negative energy are
\begin{equation}
  \label{c2}
  \begin{split}
    \phi_{3}(W) &= \psi_{1}(-W)=
    \begin{pmatrix}
      \cos{\theta(-W)}\\
      \sin{\theta(-W)}
    \end{pmatrix},\\
    \phi_{4}(W) &= \psi_{2}(-W)=
    \begin{pmatrix}
      -\sin{\theta(-W)}\\
      \cos{\theta(-W)}
    \end{pmatrix}, \\
    \psi_{3} &= \psi_{4}=0 .
  \end{split}
\end{equation}

One can write down the spectral representation of matrix $S_{1}$
\begin{equation}
  \label{c3}
  S_{1} = \lambda_{1}(W) \psi_{1} \psi_{1}^{\text{T}} +
         \lambda_{2}(W) \psi_{2} \psi_{2}^{\text{T}},
\end{equation}
where the eigenvalues $\lambda_{i}(W)$ are some functions with properties
$\lambda_{1}(m_{1})=\lambda_{2}(m_{2})=0$. So, the symmetric matrix $2 \times 2$
$S_{1}(W)$ is parametrized by three functions $\lambda_{1}(W)$, $\lambda_{2}(W)$
and $\theta(W)$. Due to the property \eqref{a15} we have
\begin{equation}
  S_{1}^{-1}=\frac{1}{\lambda_{1}(W)}\psi_{1}\psi_{1}^{\text{T}}+
  \frac{1}{\lambda_{2}(W)}\psi_{2}\psi_{2}^{\text{T}}.
\end{equation}

\section{Completeness condition and spin projectors}
\label{sec:compl-cond-spin}

The necessary requirement in constructing of spectral representation is the
completeness condition for eigenprojectors
\begin{equation}
  \label{compl}
  X \equiv \sum_{i=1}^{2n} \Pi_{i} = I_{4} I_{n} .
\end{equation}
Here $I_{4}$ and $I_{n}$ are unit matrices of indicated dimensions. If to
represent $X$ in form of decomposition \eqref{decomp} with matrix $n\times n$
coefficients $X_{M}$, then \eqref{compl} is equivalent to
\begin{equation}
  X_{1} = X_{2} = I_{n}, \qquad X_{3} = X_{4} = 0  ,
\end{equation}
or with the use of the explicit form of the projectors \eqref{L_R}:
\begin{equation}
  \label{unit}
  \begin{gathered}
    \sum_{i=1}^{2n}\psi_{i}(\tilde{\psi}_{i})^{\text{T}} =
    \sum_{i=1}^{2n}\phi_{i}(\tilde{\phi}_{i})^{\text{T}} = I_{n}, \\
    \sum_{i=1}^{2n}\psi_{i}(\tilde{\phi}_{i})^{\text{T}} =
    \sum_{i=1}^{2n}\phi_{i}(\tilde{\psi}_{i})^{\text{T}} = 0 .
  \end{gathered}
\end{equation}

Orthogonality of the projectors $\Pi_{i} \Pi_{k} = \delta_{ik} \Pi_{k}$ leads to
the property $X\cdot X = X$, i.e. $X$ may be either projector or unit
operator. To prove that sum of projectors \eqref{compl} gives the unit operator,
one should show that for arbitrary ``vector'' $\Phi$
\begin{equation}
  \label{compl1}
  X \Phi = \Phi  .
\end{equation}
\begin{itemize}
\item First of all, let us consider a single fermion field ($n=1$) in the theory
  with parity conservation. In this case $\Pi_{i} = \Lambda^{\pm}$, i.e. the
  eigenprojectors coincide with off-shell projectors \eqref{lam}. One can use
  the eigenvectors $\Pi_{i} \phi_{i}$ as basis vectors, but one needs two times
  more vectors for decomposition of arbitrary $\Phi$. Of course, the missing
  degrees of freedom are related with the spin and orthogonal basis can be
  generated by the energy projectors $\Lambda^{\pm}$ together with the spin
  projectors $\Sigma_{0}^{\pm}$, so
  \begin{multline}
    \Phi = c_{1} \Lambda^{+}(p) \Sigma_{0}^{+} (s) \phi_{1} +
      c_{2} \Lambda^{+}(p) \Sigma_{0}^{-} (s) \phi_{2} +\\
    + c_{3} \Lambda^{-}(p) \Sigma_{0}^{+} (s) \phi_{3} +
      c_{4} \Lambda^{-}(p) \Sigma_{0}^{-} (s) \phi_{4} ,
  \end{multline}
  where $\phi_{i}$ are arbitrary normalized spinors and $\Sigma_{0}^{\pm}$ are
  the standard spin projectors, commuting with $\Lambda^{\pm}(p)$:
  \begin{equation}
    \label{Sig0}
    \Sigma_{0}^{\pm}(s) = \frac{1}{2} \Big( 1 \pm \gamma^{5} \hat{s} \Big),
    \quad
    (sp)=0,\quad
    s^{2}=-1 .
  \end{equation}
  After that, the completeness condition in form of \eqref{compl1} becomes
  evident.
\item In the theory with $\mathsf{P}$-parity violation there appears a problem
  with the spin projectors. In this case the inverse dressed propagator contains
  $\gamma^{5}$ terms
  \begin{multline}
    S(p) = a(p^{2}) + \hat{n} b(p^{2}) + \gamma^{5} c(p^{2}) +
    \hat{n} \gamma^{5} d(p^{2}) ,\\
    n^{\mu}=p^{\mu}/W,\quad
    W=\sqrt{p^{2}}
  \end{multline}
  and does not commute with the standard spin projectors $\Sigma_{0}^{\pm}$. The
  eigenprojectors (solutions of the eigenvalue problem \eqref{eigenL})
  \begin{equation}
    \label{pro_1}
    \Pi_{1,2}=\dfrac{1}{2} \bigg( I_{4} \pm \hat{n}\cdot
    \dfrac{b+\hat{n}\gamma^{5} c + \gamma^{5} d}
    {\sqrt{b^{2}+c^{2}-d^{2}}} \bigg)
  \end{equation}
  also do not commute with $\Sigma_{0}^{\pm}$.

  In fact, the completeness is evident from \eqref{pro_1} since
  $\Pi_{1}+\Pi_{2}=I_{4}$, so there should exist some generalized spin
  projectors with properties
  \begin{equation}
    \label{spin_pro}
    \begin{gathered}
      \big[ \Sigma_{i}^{\pm}, \Pi_{i} \big] = 0,\quad \Sigma_{i}^{\pm}
      \Sigma_{i}^{\pm} = \Sigma_{i}^{\pm},\\
      \Sigma_{i}^{\pm}
      \Sigma_{i}^{\mp} =0,\quad \Sigma_{i}^{+} +\Sigma_{i}^{-} = I_{4} .
    \end{gathered}
  \end{equation}
  In this case the eigenvalue problem (both left and right) has twice as many
  solutions with the same orthonormality property
  \begin{equation}
    \label{spin_eig}
    S \big( \Pi_{i} \Sigma_{i}^{\pm} \big)  =
    \lambda_{i} \big( \Pi_{i} \Sigma_{i}^{\pm} \big).
  \end{equation}
  The completeness condition takes the form
  \begin{equation}
    \label{compl_s}
    \sum_{i=1}^{2} (\Pi_{i} \Sigma_{i}^{+} + \Pi_{i} \Sigma_{i}^{-}) =
    I_{4}
  \end{equation}
  and inverse propagator is represented as
  \begin{equation}
    \label{inver_s}
    S(p)=\sum_{i=1}^{2} \lambda_{i} (\Pi_{i} \Sigma_{i}^{+} +
    \Pi_{i} \Sigma_{i}^{-}) . 
  \end{equation}

  In this case ($n=1$) one can guess the answer for spin projectors. Since the
  matrices $\hat{n}$ and $\gamma^{5} \hat{s}$ have the same commutative
  properties, the spin projector is obtained from \eqref{pro_1} replacing the
  factor $\hat{n} \to \gamma^{5} \hat{s}$
  \begin{equation}
    \label{spin_n1}
    \Sigma^{\pm}=\dfrac{1}{2} \bigg(  I_{4} \pm \gamma^{5} \hat{s}\cdot
    \dfrac{b+\hat{n}\gamma^{5} c + \gamma^{5} d}
    {\sqrt{b^{2}+c^{2}-d^{2}}}
    \bigg),\quad
    s^{2}=-1,\quad (sp)=0 .
  \end{equation}
  One can easily verify that \eqref{spin_n1} have all the required properties.
  In the absence of interaction ($b=W$, $c=d=0$), or in the theory with parity
  conservation ($c=d=0$) they coincide with the standard ones
  $\Sigma_{0}^{\pm}$. So one can conclude that appearance of $\gamma^{5}$ in a
  vertex leads to dressing of spin projectors together with dressing of
  propagator.
\item With the same replacement trick $\hat{n} \to \gamma^{5} \hat{s}$ one can
  build the spin projectors in the case of $n$ fermion fields. The obtained
  eigenprojectors \eqref{L_R} may be rewritten as
  \begin{equation}
    \label{pro_n}
    \Pi_{i}= \frac{1}{2}\Big(a_{i} + \hat{n} b_{i} + \gamma^{5} c_{i} +
      \hat{n} \gamma^{5} d_{i} \Big) =
      \frac{1}{2}\Big(I_{4} I_{n} + \hat{n}t_{i} \Big),
  \end{equation}
  where $t_{i}=\hat{n}\big(a_{i} - I_{4} I_{n}\big)+ b_{i} + \hat{n} \gamma^{5}
  c_{i} + \gamma^{5} d_{i}$.

  Substitution $\hat{n} \to \gamma^{5} \hat{s}$ in last expression \eqref{pro_n}
  gives the spin projector
  \begin{equation}
    \label{spin_n}
    \Sigma_{i}= \frac{1}{2}\Big(I_{4} I_{n} + \gamma^{5} \hat{s} t_{i} \Big).
  \end{equation}

  One can check that $ \Sigma_{i}$  is actually a projector (matrices $\hat{n}$
  and $\gamma^{5} \hat{s}$ have the same properties), commuting with the
  eigenprojector $\Pi_{i}$.

  It is easy to see that $ \Sigma_{i}$ commutes with any energy projector
  $\Pi_{k}$. From \eqref{pro_n} we can express the matrix $t_{i}$
  \begin{equation*}
    t_{i}= \hat{n} \big(2 \Pi_{i} - I_{4} I_{n}\big)
  \end{equation*}
  and substitute it to the $ \Sigma_{i}$ \eqref{spin_n}
  \begin{equation}
    \label{spin_f}
    \Sigma_{i}= \frac{1}{2}\Big(I_{4} I_{n} + \gamma^{5}
                \hat{s} \hat{n} \big(2 \Pi_{i} - I_{4} I_{n}\big) \Big).
  \end{equation}
  Since the matrix $\gamma^{5} \hat{s} \hat{n}$ commutes with any
  $\gamma$-matrix in propagator
  ($I_{4}, \gamma^{5}, \hat{p}, \hat{p}\gamma^{5}$), spin projectors will
  commute with any $\Pi_{k}$
  \begin{equation}
    \big[ \Sigma_{i}, \Pi_{k} \big] = 0.
  \end{equation}
  Moreover, the expression \eqref{spin_f} is simplified essentially ``under the
  observation'' of energy projector $\Pi_{k}$ due to orthonormality property
  \begin{equation}
   \Pi_{k} \Sigma_{i}=
   \begin{cases}
     1/2 \big( I_{4}I_{n} + \gamma^{5}\hat{s}\hat{n} \big), &  k = i  \\
     1/2 \big( I_{4}I_{n} - \gamma^{5}\hat{s}\hat{n} \big), & k \neq i
   \end{cases}  
  \end{equation}
\end{itemize}

Since we have the spectral representation of the propagator \eqref{prop}, the
spin projectors $\Sigma_{i}$ are always ``under the observation'' of $\Pi_{k}$,
so the general form of spin projector (in the theory with $\gamma^{5}$) is
\begin{equation}
  \label{spin_final}
  \Sigma(s) = \frac{1}{2}\Big(I_{4} I_{n} + \gamma^{5} \hat{s} \hat{n} \Big).
\end{equation}

The existence of the spin projectors for mixing of $n$ fermion fields \eqref{spin_n}
means that we can build $4n$ eigenprojectors \eqref{spin_eig} and it proves the
completeness condition \eqref{compl}.

Let us examine the above formulas \eqref{pro_n}, \eqref{spin_n} and completeness
relation \eqref{compl} in case of two mixing fermions (see
Sec.~\ref{sec:case-parity-cons}). The projectors for positive energy poles
\eqref{a4} can be rewritten as
\begin{equation}
  \label{eq:22}
  \Pi_{i}=\psi_{i}\psi_{i}^{\text{T}}\frac{1}{2}\big( 1+\hat{n} \big)=
  \frac{1}{2}\big( 1+\hat{n}t_{i} \big),\quad
  i=1,2
\end{equation}
where
\begin{equation}
  t_{1}=\psi_{1}(\psi_{1})^{\text{T}}-\psi_{2}(\psi_{2})^{\text{T}}\hat{n},\quad
  t_{2}=\psi_{2}(\psi_{2})^{\text{T}}-\psi_{1}(\psi_{1})^{\text{T}}\hat{n},
\end{equation}
and $\psi_{1}$, $\psi_{2}$ are given by formulas \eqref{c1}. The corresponding
spin projectors are
\begin{equation}
  \label{eq:23}
  \Sigma_{i}(s)=\frac{1}{2}\big( 1+\gamma^{5}\hat{s}t_{i} \big).
\end{equation}
For this simple case the combination $\Pi_{i}\Sigma_{i}$ is simplified to
\begin{equation}
  \Pi_{i}\Sigma_{i}=\psi_{i}\psi_{i}^{\text{T}}\frac{1}{2}\big( 1+\hat{n} \big)
  \frac{1}{2}\big( 1+\gamma^{5}\hat{s} \big)=
  \psi_{i}\psi_{i}^{\text{T}}\Lambda^{+}\Sigma_{0}(s).
\end{equation}
Now it is easy to verify the completeness condition
\begin{equation}
  \label{eq:25}
  \sum_{i=1}^{4}\big( \Pi_{i}\Sigma_{i}(s)+\Pi_{i}\Sigma_{i}(-s) \big)=I_{4}I_{2}.
\end{equation}

\section{Renormalization of propagator}
\label{sec:renorm-scheme}

Let us consider the multiplicative renormalization (wave-function
renormalization) of matrix propagator $G(p)$. We restrict here ourselves by
$\mathsf{CP}$-conservating theory and by case of stable fermions.  This problem
was discussed earlier in different aspects \cite{Aoki:1982ed, Esp02, Zhou06,
  Kni12, Kni14PRL}. The main requirements for the renormalized propagator may be
found in Ref. \cite{Aoki:1982ed}, so our main purpose here is to reformulate
them in terms of the spectral representation.

If the renormalized dressed matrix propagator $G^{\text{ren}}(p)$ has poles at
points $W=\pm m_{l}$ we can put the eigenvalues $\lambda_{l}(W)$ in the same
order, so that $\lambda_{l}(m_{l}) = 0$, $l=1,\dots,n$. In vicinity of point
$\hat{p}=m_{l}$ matrix propagator has the form
\begin{equation}
  \label{eq:1}
  G^{\text{ren}}(p)\sim
  \begin{pmatrix}
    & & \vdots & & \\
    & \ldots & \dfrac{1}{\hat{p}-m_{l}} & \ldots& \\
    & & \vdots & &
  \end{pmatrix},
\end{equation}
where $(G^{\text{ren}})_{ll}$ has pole with unit residue and other elements of
$G^{\text{ren}}(p)$ are regular at $\hat{p} \to m_{l}$.  It is convenient to
renormalize the inverse matrix propagator $S(p)$, so we need to know its
behaviour in vicinity of pole. It was investigated in Ref. \cite{Aoki:1982ed},
the result may be presented in the form
\begin{equation}
  \label{inv_near}
  S^{\text{ren}}_{ij} \xrightarrow[\hat{p}\to m_{l}]{}
  \begin{cases}
    \hat{p} - m_{l}, & i=l,\, j=l,               \\
    M^{il}(\hat{p} - m_{l}), &  i\not=l,\, j=l,   \\
    (\hat{p} - m_{l})  M^{lj}, &  i=l,\, j\not=l, \\
    \text{arbitrary}, & i\not=l,\, j\not=l,
  \end{cases}
\end{equation}
where matrices $M^{il}$, $M^{lj}$ can be non-commutative with $\hat{p} - m_{l}$
because of $\gamma^{5}$. If to write down decomposition of $S^{\text{ren}}$ in
our basis
\begin{equation}
  \label{sren}
  S^{\text{ren}}(p)=\sum_{M=1}^{4} \operP_{M}\ S^{\text{ren}}_{M}(W),
\end{equation}
we can reformulate the requirements \eqref{inv_near} in terms of this
decomposition.

Note that the limit $\hat{p} \to m_{l}$ means that $p^{2}\to m_{l}^{2}$ or $W\to
\pm m_{l}$.  One can see that with use of decomposition \eqref{sren}, it's
sufficient to investigate only $W \to m_{l}$ limit (positive energy pole in
propagator) since the symmetry properties $S_{2}(W)=S_{1}(-W)$,
$S_{4}(W)=S_{3}(-W)$ guarantee the proper behaviour near the $W =-m_{l}$ point.

Let us introduce renormalization of fields in a standard manner
\begin{equation}
  \Psi= Z^{1/2} \Psi^{\text{ren}}, \quad
  \bar{\Psi} =  \bar{\Psi}^{\text{ren}} \bar{Z}^{1/2} .
\end{equation}
In theories with $\gamma^{5}$ the renormalization ``constants'' are in fact the
matrices of dimension 4
\begin{equation}
  Z^{1/2} =  \alpha + \gamma^{5} \beta, \quad
  \bar{Z}^{1/2} =   \bar{\alpha} + \gamma^{5} \bar{\beta}  .
\end{equation}
If to consider the mixing problem of $n$ generations of fermions then $\alpha$,
$\beta$, $\bar{\alpha}$, $\bar{\beta}$ are matrices of dimension $n$.

Inverse renormalized matrix propagator is defined by
\begin{equation}
  S^{\text{ren}} = \bar{Z}^{1/2} S Z^{1/2} =
  ( \bar{\alpha} + \gamma^{5} \bar{\beta}) S (\alpha + \gamma^{5} \beta).
\end{equation}

Let us restrict ourselves by $\mathsf{CP}$-conservating theory and by the case
of stable fermions. $\mathsf{CP}$-conservation leads to the symmetry properties
\eqref{sim} and in order to keep this symmetry after renormalization we have to
require\footnote{It corresponds to the pseudo-Hermitian
  condition\cite{Aoki:1982ed}
  $\bar{Z}^{1/2}=\gamma^{0} (Z^{1/2})^{\dagger} \gamma^{0}$, but in the presence
  of imaginary part in self-energy this condition becomes contradictory
  \cite{Esp02, Zhou06}.}
\begin{equation}
  \bar{\alpha} =  \alpha^{\text{T}}, \quad
  \bar{\beta} =  - \beta^{\text{T}} .
\end{equation}

So, the multiplicative renormalization of inverse propagator is defined by
\begin{equation}
  \label{renor62}
  S^{\text{ren}}(p) =  (\alpha^{\text{T}} - \gamma^{5} \beta^{\text{T}}) S(p)
  (\alpha + \gamma^{5} \beta).
\end{equation}

Renormalization conditions for $(S^{\text{ren}})_{ij}$ \eqref{inv_near} can be
formulated in terms of decomposition \eqref{sren} at
$\epsilon_{l}=W-m_{l}\to 0$.
\begin{itemize}
\item $i=l,j=l$
  \begin{equation}
    \label{cond_ll}
    \begin{split}
      (S^{\text{ren}}_{1})_{ll}& \longrightarrow W-m_{l} , \qquad
      (S^{\text{ren}}_{2}(W))_{ll} = (S^{\text{ren}}_{1}(-W))_{ll}, \\
      (S^{\text{ren}}_{3})_{ll}& = o(\epsilon_{l}) , \qquad
      (S^{\text{ren}}_{4})_{ll} = o(\epsilon_{l}) .
    \end{split}
  \end{equation}
\item $i\not=l,j=l$
  \begin{equation}
    \label{cond_il}
    (S^{\text{ren}}_{1})_{il} = O(\epsilon_{l}), \qquad
    (S^{\text{ren}}_{4})_{il} = O(\epsilon_{l}) .
  \end{equation}
  Corresponding elements of $S_{2}, S_{3}$ matrices are defined by replacement
  $W \to -W$ and they are $O(1)$.
\item $i=l,j\not=l$
  \begin{equation}
    \label{cond_lj}
    (S^{\text{ren}}_{1})_{lj} = O(\epsilon_{l}), \qquad
    (S^{\text{ren}}_{3})_{lj} = O(\epsilon_{l}) .
  \end{equation}
  Elements of matrices $S_{2}, S_{4}$ are obtained by $W \to -W$.
\end{itemize}
We see that in the limit $W\to m_{l}$ there arise some conditions on $l$-th row
and $l$-th column of $S_{1}$ matrix, on $l$-th row of $S_{3}$ and on $l$-th
column of $S_{4}$. Matrix coefficients in decomposition \eqref{sren} should have
the following behaviour at $\epsilon_{l}=W-m_{l}\to 0$
\begin{equation}
  \label{eq:2}
  \begin{aligned}
    S_{1}^{\text{ren}}&\sim
    \begin{pmatrix}
      O(1) & \ldots & O(\epsilon_{l}) & \ldots & O(1)\\
      \vdots &  & \vdots         &  &\vdots \\
      O(\epsilon_{l}) & \ldots & \epsilon_{l} & \ldots & O(\epsilon_{l}) \\
      \vdots &  & \vdots         &  &\vdots \\
      O(1) & \ldots & O(\epsilon_{l}) & \ldots & O(1),
    \end{pmatrix},\\
    S_{2}^{\text{ren}}&\sim O(1),\\
    S_{3}^{\text{ren}}&\sim
    \begin{pmatrix}
      & & O(1) & & \\
      & & \vdots & & \\
      O(\epsilon_{l}) & \ldots & o(\epsilon_{l}) & \ldots & O(\epsilon_{l})\\
      & & \vdots & & \\
      & & O(1) & &
    \end{pmatrix},\\
    S_{4}^{\text{ren}}&\sim
    \begin{pmatrix}
      & & O(\epsilon_{l}) & & \\
      & & \vdots & & \\
      O(1) & \ldots & o(\epsilon_{l}) & \ldots & O(1)\\
      & & \vdots & & \\
      & & O(\epsilon_{l}) & &
    \end{pmatrix}.
  \end{aligned}
\end{equation}

We use the spectral representation for inverse propagator \eqref{inver}, then,
according to \eqref{renor62}, the renormalized inverse propagator looks
similarly
\begin{equation}
  \label{sren_sp}
  S^{\text{ren}}=\sum_{k=1}^{2n}\lambda_{k}(W) \tilde{\Pi}_{k} ,
\end{equation}
but $\tilde{\Pi}_{k}=(\alpha^{\text{T}} - \gamma^{5} \beta^{\text{T}}) \Pi_{k}
(\alpha + \gamma^{5} \beta)$ are not projectors in general case. Recall that
projectors $\Pi_{k}$ \eqref{L_R_CP} are expressed through some vectors
${\psi}_{k}$, ${\phi}_{k}$ which we suppose to be columns.

As it turns out the operators $\tilde{\Pi}_{k}$ have the same form
\eqref{L_R_CP} with renormalized vectors
\begin{equation}
  \label{pi_tilda}
  \tilde{\Pi}_{k} = \mathcal{P}_{1}\cdot \psi_{k}^{r}
  (\psi_{k}^{r})^{\text{T}} - \mathcal{P}_{2}\cdot \phi_{k}^{r}
  (\phi_{k}^{r})^{\text{T}}
  + \mathcal{P}_{3}\cdot \psi_{k}^{r}
  (\phi_{k}^{r})^{\text{T}} - \mathcal{P}_{4}\cdot \phi_{k}^{r}
  (\psi_{k}^{r})^{\text{T}} ,
\end{equation}
where renormalized vectors look like
\begin{equation}
  \label{vectors_ren}
  \psi_{k}^{r}=\alpha^{\text{T}}\psi_{k} + \beta^{\text{T}}\phi_{k},\quad
  \phi_{k}^{r}=\alpha^{\text{T}}\phi_{k} + \beta^{\text{T}}\psi_{k}.
\end{equation}
Now require $S^{\text{ren}}$ in the form \eqref{pi_tilda} to satisfy the
conditions \eqref{eq:2}. If $W\to m_{l}$ and $\lambda_{l}(m_{l})=0$, it is
convenient to separate out the $l$-th eigenvalue in $S^{\text{ren}}$
\begin{equation}
  S^{\text{ren}}=\lambda_{l}(W) \tilde{\Pi}_{l} +
  \sum_{k\not=l}\lambda_{k}(W) \tilde{\Pi}_{k} .
\end{equation}
We will show that the renormalization conditions \eqref{eq:2} may be formulated
as requirements on the vectors $\psi_{k}^{r}(W)$. To see it, we will write the
explicit form of matrices $S^{\text{ren}}_{M}(W)$, which follows from
\eqref{sren_sp}, \eqref{pi_tilda}
\begin{equation}
  \label{SM-ren}
  \begin{split}
    S^{\text{ren}}_{1} &=
    \sum_{k}\lambda_{k}(W) \psi_{k}^{r} (\psi_{k}^{r})^{\text{T}}=
    \lambda_{l}(W) \psi_{l}^{r} (\psi_{l}^{r})^{\text{T}} +
    \sum_{k\not=l}\lambda_{k}(W) \psi_{k}^{r} (\psi_{k}^{r})^{\text{T}} , \\
    S^{\text{ren}}_{2} &=
    -\sum_{k}\lambda_{k}(W) \phi_{k}^{r} (\phi_{k}^{r})^{\text{T}}, \\
    S^{\text{ren}}_{3} &=
    \sum_{k}\lambda_{k}(W) \psi_{k}^{r} (\phi_{k}^{r})^{\text{T}}=
    \lambda_{l}(W) \psi_{l}^{r} (\phi_{l}^{r})^{\text{T}} +
    \sum_{k\not=l}\lambda_{k}(W) \psi_{k}^{r} (\phi_{k}^{r})^{\text{T}} , \\
    S^{\text{ren}}_{4} &=
    -\sum_{k}\lambda_{k}(W) \phi_{k}^{r} (\psi_{k}^{r})^{\text{T}}=
    -\lambda_{l}(W) \phi_{l}^{r} (\psi_{l}^{r})^{\text{T}} -
    \sum_{k\not=l}\lambda_{k}(W) \phi_{k}^{r} (\psi_{k}^{r})^{\text{T}} .
  \end{split}
\end{equation}

First of all, consider behaviour of the non-diagonal elements of
$S^{\text{ren}}(p)$. Looking at conditions \eqref{cond_il}, \eqref{cond_lj}, one
can see that non-diagonal elements are determined by $k\neq l$ terms in sums
\eqref{SM-ren} and are reduced to requirements on the renormalized vector
$\psi_{k}^{r}(W)$, namely
\begin{equation}
  \label{non_diag}
  (\psi_{k}^{r}(m_{l}))_{l} =0,\quad
  k\neq l .
\end{equation}

Renormalization of diagonal elements \eqref{cond_ll} is fixed by $i=l$ term in a
sum and gives the condition
\begin{equation}
  \label{diag}
  (\psi_{l}^{r}(W))_{l} \to R_{l}\neq0
  \quad \text{at } W\to m_{l} .
\end{equation}
Thus, the constant $R_{l}$ multiplying the eigenvalue, provides the unit
slope. It is naturally to suppose it as renormalized eigenvalue
\begin{equation}
  \label{eig_r}
  \lambda_{l}^{\text{ren}}(W) = \lambda_{l}(W) R_{l}^{2} \to W-m_{l}
  \quad \text{at } W\to m_{l} .
\end{equation}

Thus, the spectral representation allows to reduce the renormalization of matrix
propagator to much more simple problem \eqref{non_diag}, \eqref{diag} of
renormalization of the vectors $\psi_{k}(W)$. Solution of this problem may be
written in compact form without using perturbation theory. Let us show that
matrices $\alpha$, $\beta$ can to be chosen as
\begin{equation}
  \label{eq:renorm-matrices}
  \begin{aligned}
    \alpha &=
    \begin{pmatrix}
      R_{1}\psi_{1}(m_{1}),R_{2}\psi_{2}(m_{2}),\ldots, R_{n}\psi_{n}(m_{n})
    \end{pmatrix},\\
    \beta &= -
    \begin{pmatrix}
      R_{1}\phi_{1}(m_{1}),R_{2}\phi_{2}(m_{2}),\ldots, R_{n}\phi_{n}(m_{n})
    \end{pmatrix}.
  \end{aligned}
\end{equation}
As in the above, to simplify notations it's convenient to suppose the vectors
$\psi_{k}(W)$, $\phi_{k}(W)$, constructing the eigenprojectors $\Pi_{k}$, to be
columns. Then the matrices \eqref{eq:renorm-matrices} consist of columns ---
these vectors at fixed $W$.

Let us verify that the matrices \eqref{eq:renorm-matrices} provide the correct
renormalization properties. To this end we can calculate according to
\eqref{vectors_ren} the renormalized vector $\psi_{k}^{r}(W)$
\begin{equation}
  \label{ren_psi}
  \begin{gathered}
    {\psi}_{k}^{r}(W) =
    \begin{pmatrix}
      R_{1}\big[ \psi_{1}^{\text{T}}(m_{1})\psi_{k}(W) -
        \phi_{1}^{\text{T}}(m_{1})\phi_{k}(W) \big] \\[2ex]
      R_{2}\big[ \psi_{2}^{\text{T}}(m_{2})\psi_{k}(W) -
        \phi_{2}^{\text{T}}(m_{2})\phi_{k}(W) \big]  \\[2ex]
        \vdots  \\[2ex]
      R_{n}\big[ \psi_{n}^{\text{T}}(m_{n})\psi_{k}(W) -
        \phi_{n}^{\text{T}}(m_{n})\phi_{k}(W) \big]
    \end{pmatrix}.
  \end{gathered}
\end{equation}
Calculating the $l$-th component of this vector at the point $W=m_{l}$, we have
\begin{equation}
  ({\psi}_{k}^{r}(m_{l}))_{l}
  = R_{l}
  \bigl[ \psi_{l}^{\text{T}}(m_{l})\psi_{k}(m_{l}) -
    \phi_{l}^{\text{T}}(m_{l})\phi_{k}(m_{l}) \bigr] = R_{l} \delta_{lk},
\end{equation}
where we used the orthonormality property \eqref{ortho11}. So we see that vector
\eqref{ren_psi}, following from renormalization ``constants''
\eqref{eq:renorm-matrices} has all necessary properties and provides the correct
renormalization of inverse propagator.

\subsection{Renormalization in theory with parity conservation}
\label{sec:renorm-theory-with}

Let us illustrate the renormalization procedure by a simple example --- mixing
of two fermion fields in theory with parity conservation.

According to general recipe \eqref{eq:renorm-matrices}, in considered simple
case we have the following renormalization constant, see formulas \eqref{c1},
\eqref{c3}
\begin{equation}
  \label{c4}
  Z^{1/2}= a R =
  \begin{pmatrix}
    \psi_{1}(m_{1}),\psi_{2}(m_{2})
  \end{pmatrix} R =
  \begin{pmatrix}
    \cos{\theta(m_{1})} & -\sin{\theta(m_{2})}\\
    \sin{\theta(m_{1})} & \cos{\theta(m_{2})}
  \end{pmatrix} R,
\end{equation}
where $R=\diag(R_{1}, R_{2})$. 

Calculating the renormalized vectors \eqref{vectors_ren}, we obtain
\begin{equation}
  \label{c5}
  \psi_{1}^{r} = R a^{\text{T}} \psi_{1} = R
  \begin{pmatrix}
    \cos{\chi_{1}} \\
    \sin{\chi_{2}}
  \end{pmatrix}, \quad
  \psi_{2}^{r} = R
  \begin{pmatrix}
    -\sin{\chi_{1}} \\
    \cos{\chi_{2}}
  \end{pmatrix} ,
\end{equation}
where we introduced short notations $\chi_{1}= \theta(W)-\theta(m_{1})$,
$\chi_{2}= \theta(W)-\theta(m_{2})$.

One can write down the renormalized inverse propagator
\begin{equation}
  \label{c6}
  \begin{split}
    S_{1}^{\text{ren}} &= \lambda_{1}(W) R
    \begin{pmatrix}
      \cos^{2}{\chi_{1}} & \sin{\chi_{2}}  \cos{\chi_{1}} \\
      \sin{\chi_{2}} \cos{\chi_{1}} & \sin^{2}{\chi_{2}}
    \end{pmatrix} R + \\
    &+ \lambda_{2}(W) R
    \begin{pmatrix}
      \sin^{2}{\chi_{1}} & - \sin{\chi_{1}}  \cos{\chi_{2}} \\
      - \sin{\chi_{1}} \cos{\chi_{2}} & \cos^{2}{\chi_{2}}
    \end{pmatrix} R .
  \end{split}
\end{equation}

Renormalized propagator looks like
\begin{equation}
  \label{c7}
  \begin{split}
    G_{1}^{\text{ren}} &= \frac{1}{\lambda_{1}(W) c_{12}^{2}} R^{-1}
    \begin{pmatrix}
      \cos^{2}{\chi_{2}} & \sin{\chi_{1}}  \cos{\chi_{2}} \\
      \sin{\chi_{1}} \cos{\chi_{2}} & \sin^{2}{\chi_{1}}
    \end{pmatrix} R^{-1} + \\
    &+ \frac{1}{\lambda_{2}(W) c_{12}^{2}} R^{-1}
    \begin{pmatrix}
      \sin^{2}{\chi_{2}} & - \sin{\chi_{2}}  \cos{\chi_{1}} \\
      - \sin{\chi_{2}} \cos{\chi_{1}} & \cos^{2}{\chi_{1}}
    \end{pmatrix} R^{-1} ,
  \end{split}
\end{equation}
where $c_{12}=\cos(\theta(m_{1})-\theta(m_{2}))$.

Let us verify the behaviour of renormalized propagator at $W\to m_{1}$
\begin{equation}
  \label{c8}
  S_{1}^{\text{ren}} \to \lambda_{1}(W) R
  \begin{pmatrix}
    1     &  s_{12} \\
    s_{12} & s_{12}^{2}
  \end{pmatrix} R + \lambda_{2}(W) R
  \begin{pmatrix}
    0 & 0 \\
    0 & c_{12}^{2}
  \end{pmatrix} R,
\end{equation}
where $s_{12}=\sin(\theta(m_{1})-\theta(m_{2}))$, and
\begin{equation}
  \label{c9}
  G_{1}^{\text{ren}} \to \frac{1}{\lambda_{1}(W)} R^{-1}
  \begin{pmatrix}
    1 & 0 \\
    0 & 0
  \end{pmatrix} R^{-1} + \frac{1}{\lambda_{2}(W)} R^{-1}
  \begin{pmatrix}
    s_{12}^{2} & - s_{12} \\
    - s_{12}  & 1
  \end{pmatrix} R^{-1} .
\end{equation}

One can see, that to ensure the correct behavior \eqref{eq:1}, \eqref{inv_near}
it's enough to fix the diagonal element of matrix $R$.
\begin{equation}
  \lambda_{1}(W) R_{1}^{2} \to W-m_{1} + o(W-m_{1})  .
\end{equation}

Let us note also, that the obtained expression for renormalized propagator
\begin{equation}
  S_{1}^{\text{ren}} = \lambda_{1}(W) \psi_{1}^{r} (\psi_{1}^{r})^{\text{T}} +
                     \lambda_{2}(W) \psi_{2}^{r} (\psi_{2}^{r})^{\text{T}},
\end{equation}
is not a spectral representation of the matrix $S_{1}^{\text{ren}}$. If we
want to build the spectral representation of renormalized propagator, we need to
solve a new eigenvalue problem
\begin{equation}
  S_{1}^{\text{ren}} \Pi = \mu \Pi ,
\end{equation}
and eigenvalues $\mu_{i}(W)$ don't coincide with $\lambda_{i}(W)$ but have the
corrrect normalization properties $\mu_{i}(m_{i})=0$.

\section{Conclusions}

Here we have constructed the spectral representation for matrix fermion
propagator in the presence of $\mathsf{P}$-parity violation which gives rather
compact and simple description of the fermion mixing in the QFT. This
construction generalizes the well-known matrix spectral representation for more
complicated objects with two sets of indices.

In this representation the inverse matrix propagator has the form \eqref{inver},
where the eigenprojectors $\Pi_{i}$ are constructed \eqref{L_R} from the vectors
$\psi_{i}$, $\tilde{\psi}_{i}$. In the case of $\mathsf{CP}$-conservation we get
the simpler answer \eqref{L_R_CP} which contains only one vector $\psi_{i}$ ---
solution of homogeneous equation \eqref{cp_hom}. In this case in order to
construct the dressed propagator, we need to solve the characteristic equation
\eqref{det} for eigenvalues $\lambda_{i}(W)$ and to solve for every $i$ the
homogeneous equation \eqref{cp_hom} or \eqref{cp_phi}.

We found that the completeness condition for the projectors $\Pi_{i}$, necessary
to build the spectral representation of matrix propagator, requires to take into
account the spin degrees of freedom. The corresponding generalized spin
projectors in the theory with $\gamma^{5}$ don't coincide with the standard ones
--- see \eqref{spin_n1}, \eqref{spin_n}. When multiplied by the eigenprojectors
$\Pi_{i}$ in a propagator, they looks like universal \eqref{spin_final} for any
theory with $\gamma^{5}$ since they don't contain self-energy contributions.
But nevertheless, renormalization of $\Pi_{i}$ has also an impact on
$\Sigma_{i}$, leading to slightly different spin projectors with different $i$.

We investigated the multiplicative (WFR) renormalization of obtained matrix
propagator. The on-shell requirements of AHKKM \cite{Aoki:1982ed} for
renormalized propagator may be easily transformed into the conditions for
renormalized vector $\psi_{i}^{r}$ \eqref{non_diag}, \eqref{diag}. After that we
have much more simple problem and it allows to write down the general answer for
renormalization constants \eqref{eq:renorm-matrices}. Note that the answer for
${Z}^{1/2}$, $\bar{Z}^{1/2}$ looks very simple just in terms of vectors
$\psi_{i}(W)$ appeared in the eigenvalue problem \eqref{eigenL}.

As a result, we have an elegant algebraic construction for matrix propagator
with separated positive and negative energy poles. We suppose it will useful in
consideration of mixing and oscillation phenomena in a system of fermions.

\section*{Acknowledgments}

We are grateful to N.N. Achasov for references concerning the Regge poles and to
V.M. Leviant for reading the manuscript and useful comments.



\end{document}